\documentclass[12pt]{IEEEtran}
\usepackage{booktabs,array}
\usepackage{graphicx}
\usepackage{hyperref}
\usepackage{cite}

\begin{document}
\title{Acoustic effects of medical, cloth, and transparent face masks on
speech signals}
\author{Ryan M. Corey, Uriah Jones, and Andrew C. Singer\\
University of Illinois at Urbana-Champaign}

\maketitle

\begin{abstract}
Face masks muffle speech and make communication more difficult, especially
for people with hearing loss. This study examines the acoustic attenuation
caused by different face masks, including medical, cloth,
and transparent masks, using a head-shaped loudspeaker and a live
human talker. The results suggest that all masks attenuate frequencies
above 1 kHz, that attenuation is greatest in front of the talker,
and that there is substantial variation between mask types, especially
cloth masks with different materials and weaves. Transparent masks have poor acoustic performance compared to both medical and
cloth masks. Most masks have little effect on lapel microphones, suggesting
that existing sound reinforcement and assistive listening systems
may be effective for verbal communication with masks.
\end{abstract}

\section{Introduction}

As the world works to control the novel coronavirus 2019 (COVID-19) pandemic,
face masks are expected to prove critical to slowing the spread of
the virus. However, it can be difficult to understand
speech when the talker is wearing a mask, especially for listeners
with hearing loss \cite{chodosh2020face,tucci2020cloth}.
By studying the acoustic effects of masks on speech signals, we can
determine which masks are best for speech
transmission and evaluate technologies to make communication easier.

Most prior research on masked speech has focused on medical equipment
such as surgical masks and N95 respirators. A recent study on the
acoustics of medical masks showed that surgical masks and N95 respirators
can attenuate higher-frequency sounds by between 3 and 12 dB \cite{goldin2020masks}.
Listening tests using audio-only recordings made with medical masks
have not shown significant effects on speech intelligibility \cite{mendel2008speech,thomas2011does,palmiero2016speech}. 

To conserve supplies of medical masks, health authorities have recommended
cloth masks, which can be made from household materials or purchased
commercially. Recent studies suggest that the effectiveness of cloth
masks at blocking respiratory droplets depends on the fabric material,
weave, and thickness \cite{Aydin2020fabrics,konda2020aerosol}. Because
both medical and cloth face masks obstruct visual cues that contribute to speech intelligibility \cite{llamas2009effects}, some hearing loss advocates
recommend the use of transparent face coverings \cite{tucci2020cloth}.
In listening tests with audiovisual recordings of talkers wearing lapel microphones, masks with clear windows were shown to improve intelligibility for listeners
with severe-to-profound hearing loss compared to paper masks \cite{atcherson2017effect}.

\begin{figure}
\begin{centering}
\includegraphics[width=8cm]{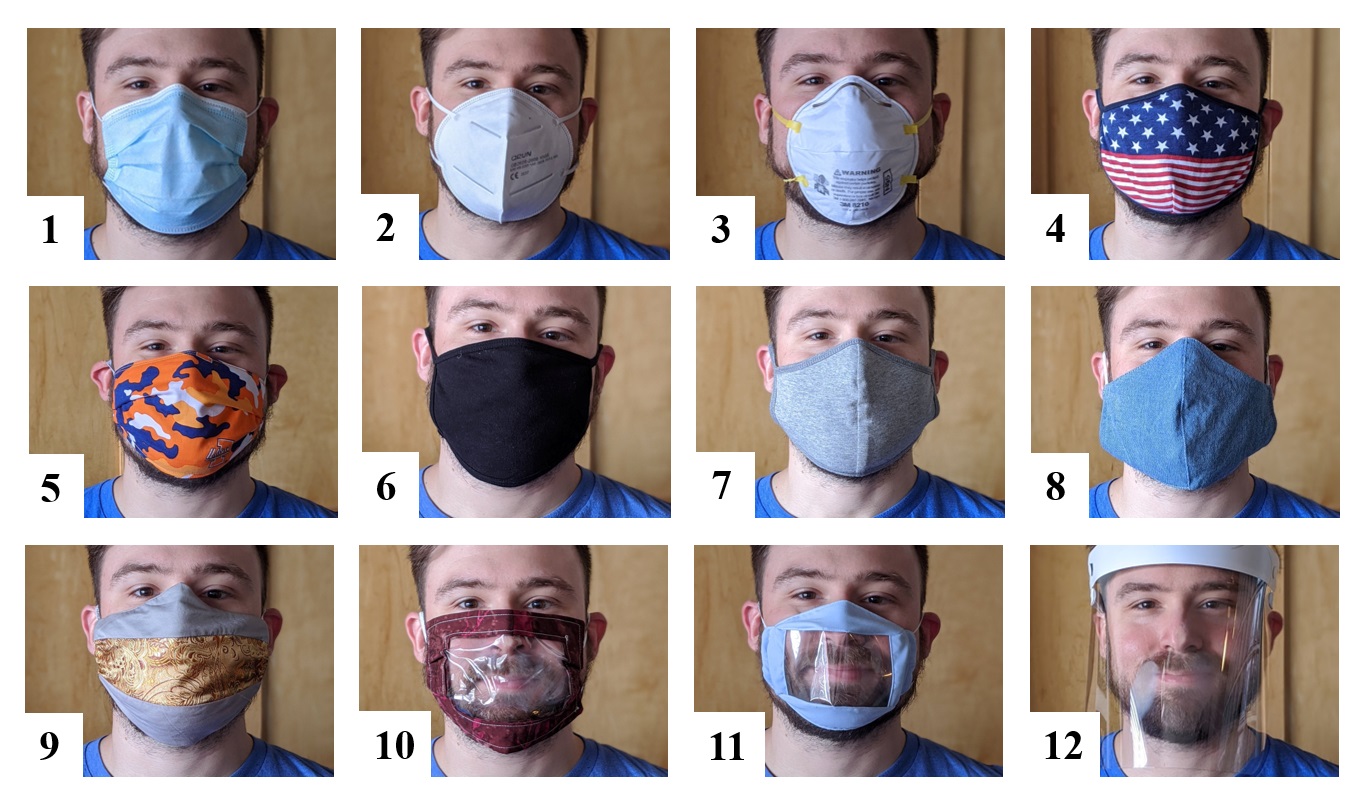}
\par\end{centering}
\caption{\label{fig:photos}Masks used in experiments and described in \autoref{tab:Face-coverings}}
\end{figure}

To understand the effects of masks on speech, we measured
the acoustic attenuation of a polypropylene
surgical mask, N95 and KN95 respirators, six cloth masks made from
different fabrics, two cloth masks with transparent windows, and a plastic shield, as shown in \autoref{fig:photos}. Measurements
were performed using both a
head-shaped loudspeaker and a live human talker. The experiments show
that different masks have different high-frequency effects
and that they alter the directivity of speech. Finally, to examine
the effects of masks on sound reinforcement and assistive listening
systems, we took measurements with microphones placed on the lapel,
cheek, forehead, and next to the mouth. These amplification technologies
may prove critical to verbal communication during the pandemic.

\section{Methods}

\begin{figure}
\begin{centering}
\includegraphics[width=8cm]{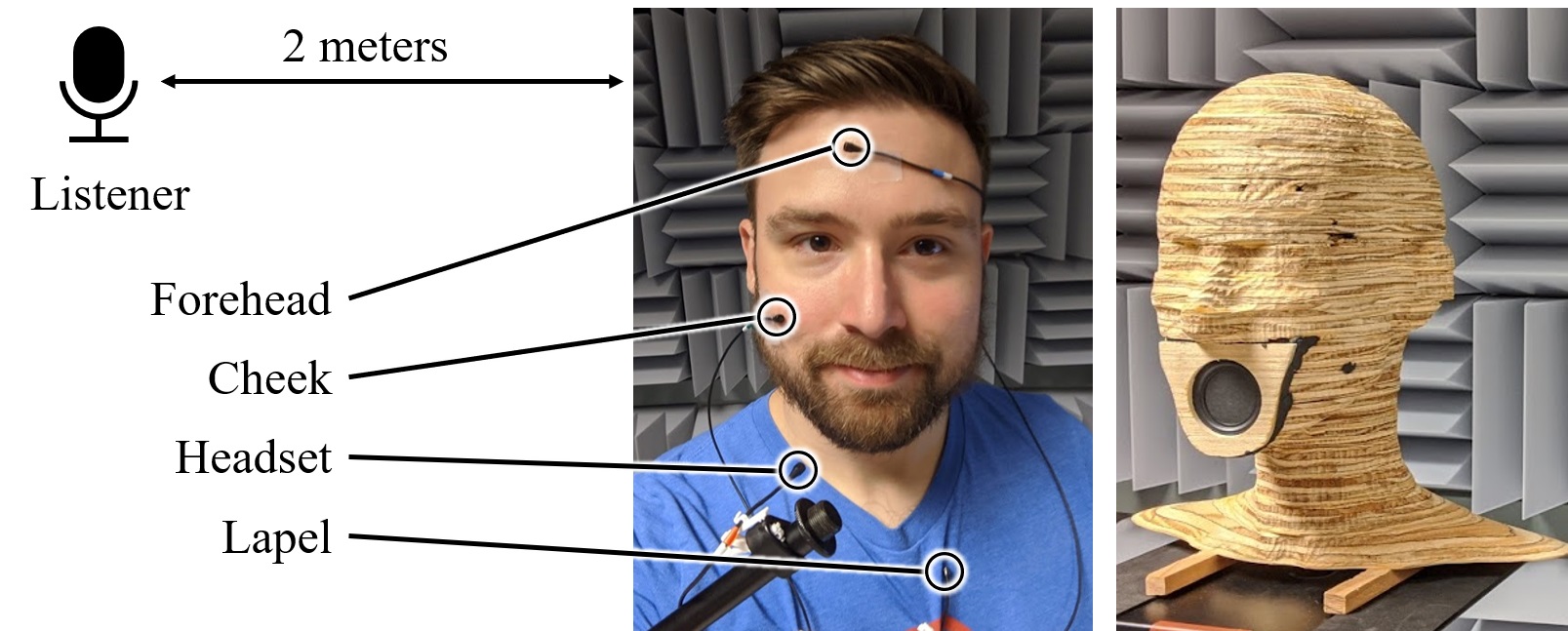}
\par\end{centering}
\caption{\label{fig:setup}Speech signals were produced by a human talker and loudspeaker
model. Microphones were placed
at listener distance and at several points on and near the face.}
\end{figure}

To simulate sound heard by a conversation partner, a side-address
cardioid condenser microphone was placed two meters from the talker position. To study the effect of
masks on sound reinforcement and assistive listening systems, omnidirectional
lavalier condenser microphones were placed next to the mouth (``headset''
position), on the lapel, on the cheek, and on
the forehead of the talker, as shown in \autoref{fig:setup}. The
laboratory walls are acoustically treated with 8-inch melamine and 2-inch polyurethane foam wedges.

Sound was produced by two sources. A custom-built head-shaped loudspeaker
produced ten-second logarithmic frequency sweeps to measure acoustic
transfer functions between the talker and listener positions. The
plywood loudspeaker uses a 2-inch full-range driver and has a directivity
pattern that is closer to that of a human talker compared to
studio monitors. To characterize the directional effects of masks,
the loudspeaker was placed on a turntable and rotated in 15 degree
increments while the ``listener'' microphone remained fixed.

For more realistic speech signals, 30-second read-speech recordings
were made from a human talker, who attempted
to use a consistent speech level for each recording. Recordings
of the human talker were repeated three times non-consecutively with
each mask. Human subject research was approved by the University of
Illinois Institutional Review Board with protocol number 19503.

\begin{table*}
\caption{\label{tab:Face-coverings}Mask measurements and 2--16 kHz acoustic attenuation results}
\centering
\begin{tabular}{>{\raggedleft\arraybackslash}p{0.5cm}<{$\,\,$} p{3.8cm} >{\raggedleft\arraybackslash}p{1.4cm} >{\raggedleft\arraybackslash}p{1.4cm} >{\raggedleft\arraybackslash}p{1.4cm} >{\raggedleft\arraybackslash}p{2.3cm} >{\raggedleft\arraybackslash}p{2.3cm} >{\raggedleft\arraybackslash}p{2.3cm}}
\toprule
 & Material & Layers & Thickness (mm) & Mass (g) & Speaker atten. at listener (dB) & Human atten. at listener (dB) & Human atten. at lapel (dB) \\
\midrule
1 & Polypropylene surgical & 3 & 0.4 & 3 & 3.6 & 2.8 & 1.0\\
2 & KN95 respirator (GB2626) & 2 & 0.6 & 4 & 4.0 & 2.6 & 0.0\\
3 & N95 respirator (3M 8210) & 1 & 1.5 & 9 & 5.7 & 5.4 & 3.6\\
4 & Cotton jersey & 2 & 0.7 & 11 & 4.0 & 3.1 & 0.5\\
5 & Cotton plain & 2 & 0.5 & 11 & 4.0 & 4.3 & 1.4\\
6 & Cotton/spandex jersey & 3 & 1.5 & 16 & 6.1 & 5.2 & 2.3\\
7 & Cotton/spandex jersey & 2 & 0.9 & 17 & 8.2 & 6.1 & 2.0\\
8 & Cotton plain \& denim & 2 & 1.1 & 21 & 9.4 & 10.0 & 3.2\\
9 & Cotton percale bedsheet \& polyester trim & 2 & 1.0 & 14 & 12.6 & 9.5 & 3.1\\
10 & Cloth \& vinyl window & 1 & 0.4 & 12 & 10.8 & 7.8 & $-2.0$\\
11 & Cloth \& PVC window & 1 & 0.3 & 7 & 12.5 & 8.0 & 0.4\\
12 & Plastic shield & 1 & 0.4 & 50 & 13.7 & 8.2 & $-7.6$\\
\bottomrule
\end{tabular}
\end{table*}

For both the loudspeaker and human experiments, measurements were
first taken with no face covering to establish a baseline. The recordings
were then repeated with the twelve face coverings listed in
\autoref{tab:Face-coverings} and shown in \autoref{fig:photos}. 

\section{Results and Discussion}

\subsection{Acoustic attenuation of face coverings}

\begin{figure*}
\begin{centering}
\includegraphics{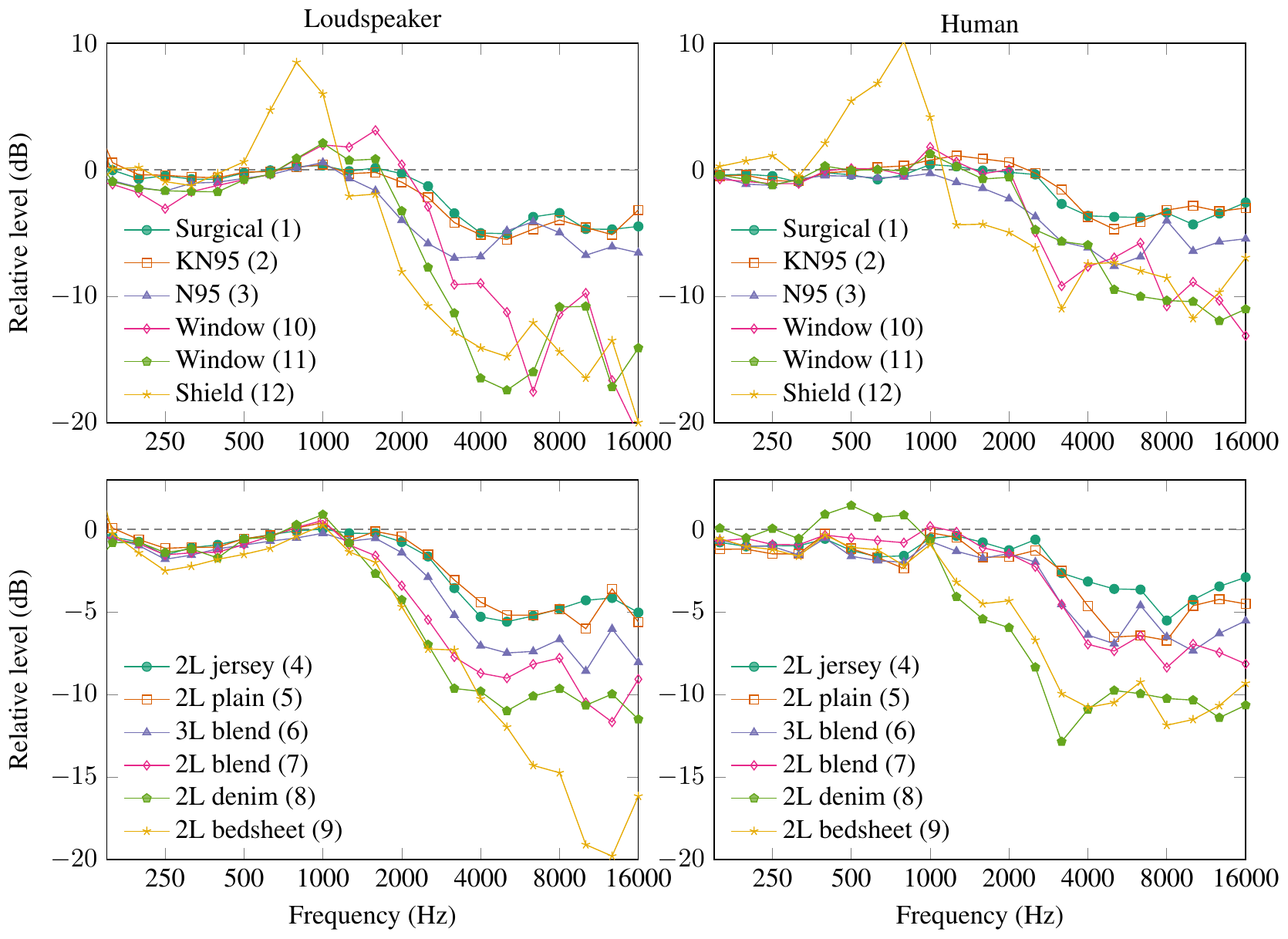}
\par\end{centering}
\caption{\label{fig:attenuation}Effect of different masks on sound levels
measured at the listener position for a head-shaped loudspeaker (left)
and human talker (right).}
\end{figure*}

\autoref{fig:attenuation} shows the effects of several
masks measured at the listener position. The plots on the left show
the differences in acoustic transfer functions measured with and without
masks on the head-shaped loudspeaker. The plots on the right show
the corresponding results for the human talker averaged over three
non-consecutive recordings; the human spectra varied by roughly 1 dB between
recordings. The attenuation values shown in Table \ref{tab:Face-coverings}
are logarithmically weighted averages from 2 kHz to 16 kHz, that is,
means of the points shown in the plots.

Most masks had little effect below 1 kHz but they
attenuated higher frequencies by different amounts. The surgical mask (1) and
KN95 respirator (2) had peak attenuation of around 4 dB, which is consistent
with the results reported by Goldin et al. \cite{goldin2020masks} with a head-and-torso
simulator. The N95 respirator (3) attenuated high frequencies by about
6 dB, which is similar to the average attenuation reported by Goldin et al.
\cite{goldin2020masks}.

The cloth masks varied widely depending on composition and weave.
The 100\% cotton masks in jersey (4) and plain (5) weaves had the
best acoustic performance and were comparable to the surgical mask.
The cotton/spandex blends performed worse. Surprisingly, the 2-layer
cotton/spandex mask (7) produced greater attenuation than the 3-layer
cotton/spandex mask (6), perhaps because it has a higher proportion of
spandex and fit more snugly on the face. Masks made from tightly woven denim (8) and bedsheets (9) performed worst acoustically. It appears
that material and weave are the most important variables determining
the acoustic effects of cloth face masks: More breathable fabrics
transmit more sound.

Finally, the transparent masks (10--12) performed poorly acoustically at high frequencies,
blocking around 8 dB for the human talker and 10--14 dB for the loudspeaker.
Although these masks are often recommended to help listeners with
hearing loss because they preserve visual cues, they also harm the
high-frequency sound cues that are crucial for speech.

\subsection{Effect of face coverings on speech directivity}

\begin{figure}
\begin{centering}
\includegraphics{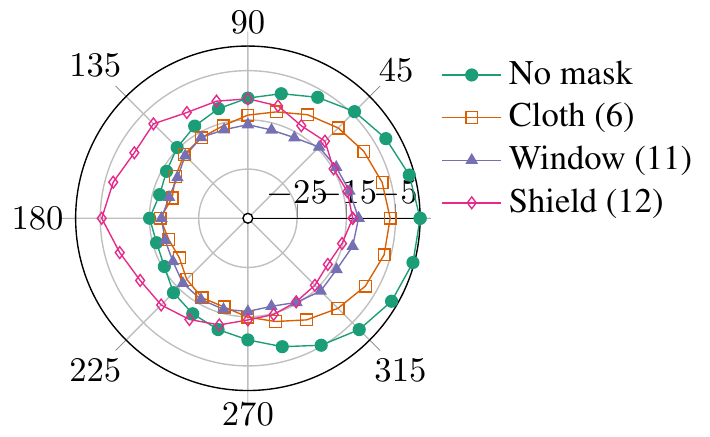}
\par\end{centering}
\caption{\label{fig:directivity}Spatial distribution of 2--16 kHz sound energy
for a head-shaped loudspeaker with different masks, in dB relative
to no mask at 0 degrees.}
\end{figure}

\autoref{fig:directivity} shows the relative
high-frequency sound level as a function of angle for the head-shaped
loudspeaker. The plot shows a logarithmically weighted average of
relative sound level from 2 kHz to 16 kHz. For all masks tested, acoustic
attenuation was strongest in the front. Sound transmission to the
side of and behind the talker was less strongly affected by the masks, and the shield (12) amplified sound behind the talker.
These results suggest that masks may deflect sound energy to the sides
rather than absorbing it. Therefore, it may be possible to use microphones
placed to the side of the mask for sound reinforcement.

\subsection{Effect of microphone placement}

\begin{figure}
\begin{centering}
\includegraphics{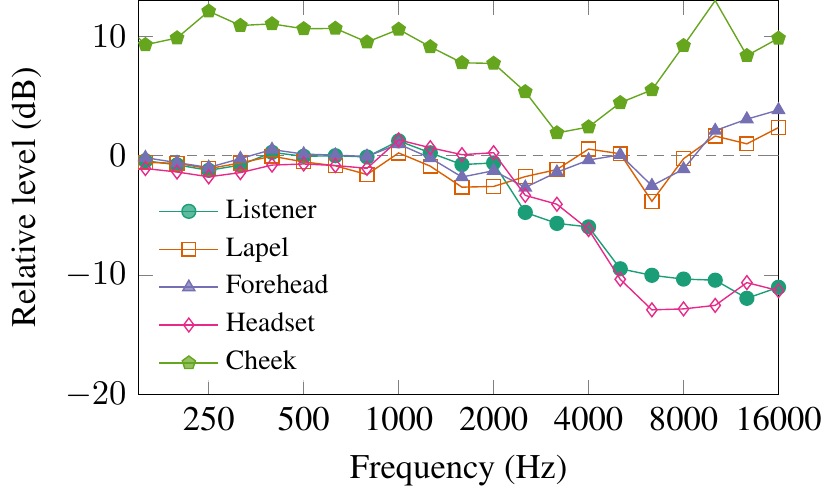}
\par\end{centering}
\caption{\label{fig:mics}Effect of mask 11 on sound levels measured at different
microphones relative to the same measurements with no mask on a human
talker.}
\end{figure}

\begin{figure}
\begin{centering}
\includegraphics{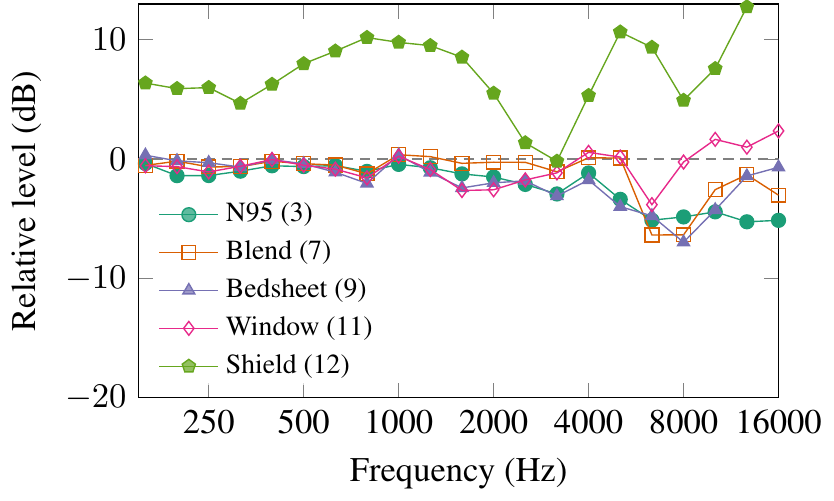}
\par\end{centering}
\caption{\label{fig:lapel}Effect of several masks on sound levels at the lapel microphone on a human talker, relative to the same measurements with no mask.}
\end{figure}

Masks attenuate high-frequency sound for distant listeners, but they have different effects on microphones on and near
the face. \autoref{fig:mics} shows the acoustic effects of the
PVC window mask (11) on different microphones on
a human talker. The listener and headset microphones experience similar
high-frequency attenuation. The cheek microphone
taped under the mask recorded higher sound levels, but with spectral
distortion. The lapel and forehead microphones showed small and
mostly uniform attenuation over the range of speech frequencies. Similar
results were obtained for masks 1--10, although the performance
of the cheek microphone varied depending on the shape of the mask.
The shield (12) strongly distorted speech spectra for all microphones.

\autoref{fig:lapel} compares the performance of several masks using a lapel microphone. 
Only the shield has a substantial effect on the speech spectra captured by the microphone.
Sound capture and reinforcement systems used in classrooms and lecture
halls often rely on lapel microphones, and remote microphones that transmit to hearing
aids are also often worn on the chest. These systems should work with most masks
with little modification. It is worth noting that lapel microphones were used for the
audiovisual recordings of \cite{atcherson2017effect},
which showed intelligibility benefits with clear masks.

\section{Conclusions}

The experimental results presented here confirm that face masks attenuate
high-frequency sound in front of the talker, with the strongest attenuation
above 4 kHz. Ubiquitous polypropylene surgical masks offer the best
acoustic performance among all masks tested. If those masks are not
available, loosely woven 100\% cotton masks perform well acoustically.
Tightly woven cotton and cotton/spandex blends should be avoided if
speech transmission is a concern. It is important to note that this
study did not consider the efficacy of masks at blocking respiratory droplets;
it is possible that loosely woven fabrics that perform well acoustically
are less effective against the virus and vice versa.

Shields and masks with windows perform much worse acoustically than opaque
cloth masks. Fortunately, window masks do not strongly affect the lapel microphones
used in sound reinforcement and assistive listening systems. To preserve
visual cues without destroying high-frequency sound cues,
talkers can wear clear window masks and lapel microphones. Although
face masks make verbal communication more difficult, amplification
technologies can help people with and without hearing loss to communicate
more effectively during the pandemic.

\section*{Acknolwedgments}
Mask 5 was sewn by Ms. Catherine Somers
and mask 9 was sewn by Mr. Austin Lewis.
This research was supported by an appointment to the Intelligence
Community Postdoctoral Research Fellowship Program at the University
of Illinois at Urbana-Champaign, administered by Oak Ridge Institute
for Science and Education through an interagency agreement between
the U.S. Department of Energy and the Office of the Director of National
Intelligence.


\end{document}